\documentclass[onecolumn,aps,preprint,showpacs,amsmath,amssymb]{revtex4}
\usepackage{graphicx}
\usepackage{latexsym}
\usepackage{dcolumn} 
\usepackage{stmaryrd} 
 
\usepackage{bm}

\begin{document}
\newcommand{\eq}{\begin{equation}}                                                                         
\newcommand{\eqe}{\end{equation}}             
\newcommand{\pa}{\partial}

\title{Analytic self-similar solutions of  the Oberbeck-Boussinesq equations} 

\author{ I.F. Barna$^{1,2}$ and L. M\'aty\'as$^3$ }
\address{ $^1$ Wigner Research Center of the Hungarian Academy of Sciences 
\\ Konkoly-Thege \'ut 29 - 33, 1121 Budapest, Hungary \\
$^2$  ELI-HU Nonprofit Kft.,  Dugonics T\'er 13, 6720 Szeged, Hungary \\ 
$^3$ Sapientia University, Faculty of Science, Libert\u{a}tii sq. 1, 530104 Miercurea Ciuc, Romania}

\date{\today}

\begin{abstract} 

In this article we will present pure two-dimensional analytic  solutions for the coupled non-compressible Newtoniain Navier-Stokes
 --- with Boussinesq approximation --- and the heat conduction equation. The system was investigated from E.N. Lorenz half a century ago with Fourier series and pioneered the way to the paradigm of chaos.  We present a novel analysis of the same system where the key idea is the two-dimensional generalization of the well-known self-similar Ansatz of Barenblatt which will be interpreted in a geometrical way. The results, the pressure, temperature and velocity fields are all analytic and can be expressed with the help of the error functions.  
The temperature field has a strongly damped oscillating behavior which is an interesting feature.  
\end{abstract}

\pacs{47.10.ad,02.30.Jr}
\maketitle

\section{introduction}

The investigation of the dynamics of viscous fluids has a long past. 
Enormous scientific literature is available from the last two centuries for fluid motion even without any kind of heat exchange. 
Thanks to new exotic materials like nanotubes,  heat conduction in solid bulk phase (without any kind of material transport) is an other quickly growing independent research area as well.  The combination of both processes are even more complex which lacks  general existence theorems for unique solutions. 
The most simple way to couple these two phenomena together is the Boussinesq \cite{bous} approximation which is used in the field of buoyancy-driven flow (also known as natural convection). 
It states that density differences are sufficiently small to be neglected, except where they appear in terms multiplied by g, the acceleration due to gravity. The main idea of the Boussinesq approximation is that the difference in inertia is negligible but gravity is sufficiently strong to make the specific weight appreciably different between the two fluids. When the Boussinesq approximation is used than no sound wave can be described in the fluid, because sound waves move via density variation. 

Boussinesq flows are quite common in nature (such as  oceanic circulations, atmospheric fronts or  katabatic winds), industry (fume cupboard ventilation or dense gas dispersion), and the built environment (like central heating, natural ventilation). The approximation is extremely accurate for  such flows, and makes the mathematics and physics much simpler and transparent. 

The advantage of the approximation arises because when investigation a flow of, say, warm and cold waters with densities $\rho_1$ and  $\rho_2$ are considered,  
 the difference $\Delta \rho = \rho_1 -  \rho_2$  is negligible and one needs only a single density $\rho$. It can be shown with the help of dimensional analysis, under these circumstances, the only sensible way that acceleration due to gravity g should enter into the equations of motion is in the reduced gravity  $g' = g (\rho_1 -\rho_2)$. The corresponding dimensionless numbers for such flows are the Richardson and Rayleigh numbers.   
The used mathematics is therefore much simpler because the density ratio ($\rho_1/\rho_2$ a dimensionless number) is exactly one and  does not affect the features of the investigated flow system. 

In the following we analyze the dynamics of a two-dimensional viscous fluid with additional 
heat conduction mechanism.  
Such systems were first investigated by Boussinesq \cite{bous} and Oberbeck \cite{ober} in the nineteenth century.  Oberbeck used a finite series expansion.  He developed a model to study the heat convection in fluids 
taking into account the flow of the fluid as a result of temperature difference.
He applied the model to the normal atmosphere.  

More than half a century later Saltzman \cite{salz} tried to solve the same model with the help 
of Fourier series.  At the same time Lorenz \cite{lorenz} analyzed the solutions with computers 
and published the plot of a strange attractor which was a pioneering results and the advent of the studies of chaotic dynamical systems. The literature of chaotic dynamics is enormous but a modern basic introduction can be found in \cite{tamas}.    

Later till to the first  beginning years of the millennium \cite{lorenz} Lorenz analyzed the final first order chaotic ordinary differential equation(ODE) system with different 
numerical methods.  This ODE system becomes an emblematic object of chaotic systems  and 
attracts much interest till today  \cite{lain}.  

On the other side critical studies came to light which go beyond the simplest truncated Fourier series. Curry for example gives a transparent proof that the finite dimensional approximations have bounded solutions \cite{curry}. Musielak {\it{et al}} \cite{musi} in three papers analyzed large number of truncated systems with different kinds and found chaotic and periodic solutions as well. 
 The messages of these studies will be shortly mentioned  later.   

In our study  we apply a completely different investigation approach, namely the two-dimensional generalization of the self-similar Ansatz  which is well-known for one dimension from  more than half a century \cite{sedov,barenb,zeld}.  This generalized Ansatz was successfully applied to the three dimensional compressible and non-compressible Navier-Stokes equations \cite{imre2, imre4} from us in the last years. 
We investigated one dimensional Euler equations with heat conduction as well \cite{imre3}  which can be understood as the precursor of the recent study.   

To our knowledge this kind of investigation method was not yet applied to the Oberbeck-Boussinesq (OB) system. 
In the next section we outline our theoretical investigation together with the results. 
The paper ends with a short summary.   

\section{Theory and Results} 

We consider the original partial differential equation(PDE) system of Saltzman \cite{salz} to 
describe heat conduction in a two dimensional viscous incompressible fluid.  
In Cartesian coordinates and Eulerian description  these equations 
have the following form:  
\begin{eqnarray}
\frac{\partial u}{\partial t} + u \frac{\partial u}{\partial x} + w \frac{\pa u}{\pa z} + 
\frac{\pa P}{\pa x} - \nu \left( \frac{\pa^2 u} {\pa x^2} + \frac{\pa^2 u} {\pa z^2} \right) = 0, \nonumber \\ 
\frac{\partial w}{\partial t} + u \frac{\partial w}{\partial x} + w \frac{\pa w}{\pa z} + 
\frac{\pa P}{\pa z} - eG T_1 - \nu \left( \frac{\pa^2 w} {\pa x^2} + \frac{\pa^2 w} {\pa z^2} \right) = 0, \nonumber \\ 
\frac{\pa T_1}{\pa t} + u \frac{\pa T_1}{\pa x}  + w \frac{\pa T_1}{\pa z} 
- \kappa \left( \frac{\pa^2 T_1 } {\pa x^2} + \frac{\pa^2 T_1} {\pa z^2} \right) =0,  \nonumber \\ 
\frac{\pa u}{\pa x} + \frac{\pa w}{\pa z} = 0,                 
\label{nav} 
\end{eqnarray}
 where $u,w,  $ denote respectively the x and z velocity coordinates, $T_1$ is the temperature difference relative
to the average ($T_1 = T - T_{av}$) and $P$ is the scaled pressure over the density .
The free physical parameters are $\nu,  e, G, \kappa $ kinematic viscosity, coefficient of volume expansion,  
acceleration of gravitation and coefficient of thermal diffusivity. 
 (To avoid further misunderstanding we use $G$ for gravitation acceleration and  g which is reserved for a self-similar solution.)\
   The first two equations are the Navier-Stokes equations, the  third one is the heat conduction equation and the last one is the continuity equation all are for two spatial dimensions. The Boussinesq approximation means the way how the heat conduction is 
coupled to the second NS equation.
Chandrasekhar \cite{chandr} presented a wide-ranging discussion of the physics and mathematics of Rayleigh-Benard convection along with many historical references.

Every two dimensional flow problem can be reformulated with the help of the stream function $\Psi$ via $u = \Psi_y$ and $v = -\Psi_x$ 
 which automatically fulfills the continuity equation.  The subscripts mean partial derivations. 
After introducing dimensionless quantities the system of  (\ref{nav}) is reduced to the next two PDEs 
 \begin{eqnarray} \nonumber  \\ 
(\Psi_{xx} + \Psi_{yy})_t + \Psi_x(\Psi_{xxz} + \Psi_{yyz})  
- \Psi_z(\Psi_{xxx} + \Psi_{zzx})- \nonumber \\ \sigma(\theta_x - \Psi_{xxxx}- \Psi_{zzzz} - 2\Psi_{xxzz}) = 0, \nonumber \\   
\theta_t + \Psi_x\theta_z - \Psi_z\theta_x
 - R \Psi_x - ( \theta_{xx} + \theta_{zz}) = 0, 
\label{stream}
\end{eqnarray}
where $\Theta$ is the scaled temperature, $\sigma = \nu/\kappa$ is the Prandtl Number 
and $R = \frac{GeH^3 \Delta T_0}{\kappa\nu}$ is the Rayleigh number  and H is the height of the fluid. 
A detailed derivation of (\ref{stream}) can be found in \cite{salz}.

All the mentioned studies in the introduction, investigated these two PDEs with the help of some 
truncated Fourier series, different kind of truncations are available which result different 
ordinary differential equation(ODE) systems. 
The derivation of the final non-linear ODE system from the PDE system can be found in the original papers \cite{salz,lorenz}.  Berg\'{e} {\it{et al.}} \cite{berge} contains a slightly different development of the Lorenz model equatios, and in addition, provides more details on how the  dynamics evolve as the reduced Rayleigh number changes.  
 The book of Sparrow \cite{spar} gives a detailed treatment of the Lorenz model and its behavior as well. Hilborn \cite{hilb} presents the idea of the derivation in a transparent and easy way. 
Therefore, we do not mention this derivation in our manuscript. 

 Some truncations violates energy conservation \cite{lain} and some not. Roy  and Musiliak \cite{musi} in his exhausting three papers present various energy-conserving truncations. Some of them contain horizontal modes, some of them contain vertical modes and some of them both kind of modes in the truncations. 
All these models show different features some of them are chaotic and some of them  - in well-defined parameter regimes - show periodic orbits in the projections of the phase space.  This is a  
true indication of the complex nature of the original flow problem. It is also clear that the Fourier expansion method which is a two hundred year old routine tool for linear PDEs fails for a relevant non-linear PDE system. 

Therefore, we apply another investigation method which is common for non-linear PDEs. 
At first we introduce the two dimensional generalization of the self-similar Ansatz

\eq 
v(x,t)=t^{-\alpha}f\left(\frac{x}{t^\beta}\right):=t^{-\alpha}f(\eta)
\label{self}
\eqe 
where $v(x,t)$ can be an arbitrary variable of a PDE and $t$ means time and $x$ means spatial 
dependence.
The similarity exponents $\alpha$ and $\beta$ are of primary physical importance since $\alpha$  represents the rate of decay of the magnitude $v(x,t)$, while $\beta$  is the rate of spread 
(or contraction if  $\beta<0$ ) of the space distribution for $t > 0 $.
The most powerful result of this Ansatz is the fundamental or 
Gaussian solution of the Fourier heat conduction equation (or for Fick's
diffusion equation) with $\alpha =\beta = 1/2$. These solutions are exhibited on Figure 1. for time-points $t_1<t_2$. 
This transformation
is based on the assumption that a self-similar solution
exists, i.e., every physical parameter preserves its
shape during the expansion. Self-similar solutions usually
describe the asymptotic behavior of an unbounded or a far-field
problem; the time t and the space coordinate x appear
only in the combination of  $f(x/t^{\beta})$. It means that the existence
of self-similar variables implies the lack of characteristic
lengths and times. These solutions are usually not unique and
do not take into account the initial stage of the physical expansion process.
It is also transparent from (\ref{self}) that to avoid singularity at $t=0$  the following 
transformation $\tilde{t} = t+t_0$ is valid. 

There is a reasonable generalization of (\ref{self})  in the form of $v(x,t) = h(t)\cdot f[x/g(t)] $, where $h(t), g(t)$ are continuous functions. The choice of  $h(t)=g(t) = \sqrt{t_0-t}$  is 
called the blow-up solution, which means that the solution becomes infinity after a well-defined finite time duration. 

These kind of solutions  describe the intermediate asymptotic of a problem: they hold when the precise initial conditions are no longer important, but before the system has reached its final steady state. For some systems it can be shown that the self-similar solution fulfills the source type (Dirac delta) initial condition. 
They are much simpler than the full solutions and so easier to understand and study in different regions of parameter space. A final reason for studying them is that they are solutions of a system of ODEs and hence do not suffer the extra inherent numerical 
problems of the full PDEs. In some cases self-similar solutions helps to understand diffusion-like properties or the existence of compact supports of the solution. 

Let's introduce the two dimensional generalization of the self-similar Ansatz
(\ref{self}) which might have the general form of 
\eq
v(x,z,t) = t^{-\alpha} f\left(  \frac{F(x,z)}{t^{\beta}} \right)
	\eqe 
where $F(x,z)$ could be understood as an implicit parametrization of a one-dimensional space curve with continuous first and second derivatives. 
 In our former studies \cite{imre2,imre4} we explain in heavy details that for the Navier-Stokes type of non-linearity  unfortunately only the $ F(x,z) = x+z+c$ function is valid in Cartesian coordinates which is a straight line. 
It basically comes from the symmetry properties of the left-hand of the  NS equation. 
Only this function fulfills the following relation  $u_x = u_z$. 
(Other locally orthogonal coordinate systems e.q. spherical are not investigated yet.) 

We may investigate both dynamical systems, the original hydrodynamical (\ref{nav}) or the other one (\ref{stream}) which is valid for the stream functions. 
 
Similar to the former studies  \cite{salz,lorenz} try to solve the PDEs for the dimensionless stream and temperature functions in the form of   
\begin{eqnarray}
\Psi = t^{-\alpha}f(\eta),    \hspace*{1mm} \theta = t^{-\epsilon}h(\eta), \hspace*{1mm} \eta = \frac{x+z}{t^{\beta}}. 
\label{anss}
\end{eqnarray}
 
Unfortunately, after some algebra it becomes clear that the constraints which should fix the values of the exponents become contradictory, therefore no unambiguous ODE can be derived.     
This means that the PDE of the stream function and the dimensionless temperature do not have self-similar solutions. In other words these functions have no such a diffusive property which 
could be investigated with the self-similar Ansats, which is a very instructive example of the applicability of the trial function of (\ref{anss}). Our experience shows that, most of the 
investigated PDEs have a self-similar ODE system and this is a remarkable exception. 

Now investigate the original hydrodynamical system with the next Ansatz 
\begin{eqnarray}
u(\eta) = t^{-\alpha} f(\eta),  \hspace*{2mm} 
w(\eta) = t^{-\delta} g(\eta), \hspace*{2mm}
P(\eta) = t^{-\epsilon} h(\eta), \hspace*{2mm} 
T_1(\eta) = t^{-\omega} l(\eta),  
\label{ans}
\end{eqnarray}

\begin{figure} 
\scalebox{0.4}{
\rotatebox{0}{\includegraphics{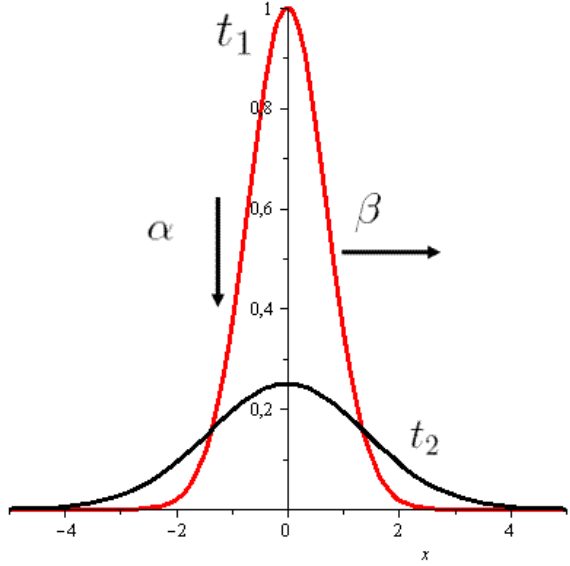}}}
\caption{A self-similar solution of Eq. (\ref{self}) for $t_1<t_2$.
The presented curves are Gaussians for regular heat conduction.}	
\label{egyes}       
\end{figure}

\begin{figure} 
\scalebox{0.45}{
\rotatebox{0}{\includegraphics{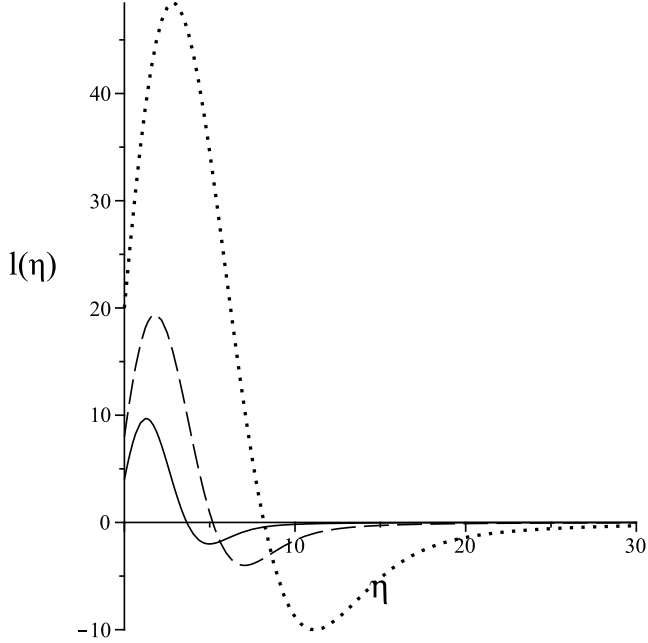}}}
\caption{Different shape functions of the temperature Eq. (\ref{temp}) as a function of $\eta$
for different thermal diffusivity. The integration constants are   $c_1 =c_2 =1$ the same for all the three curves. The solid the dashed and the dotted lines are for $\kappa =1,2,5$, respectively.}	
\label{egyes}       
\end{figure}
\begin{figure} 
\scalebox{0.45}{
\rotatebox{0}{\includegraphics{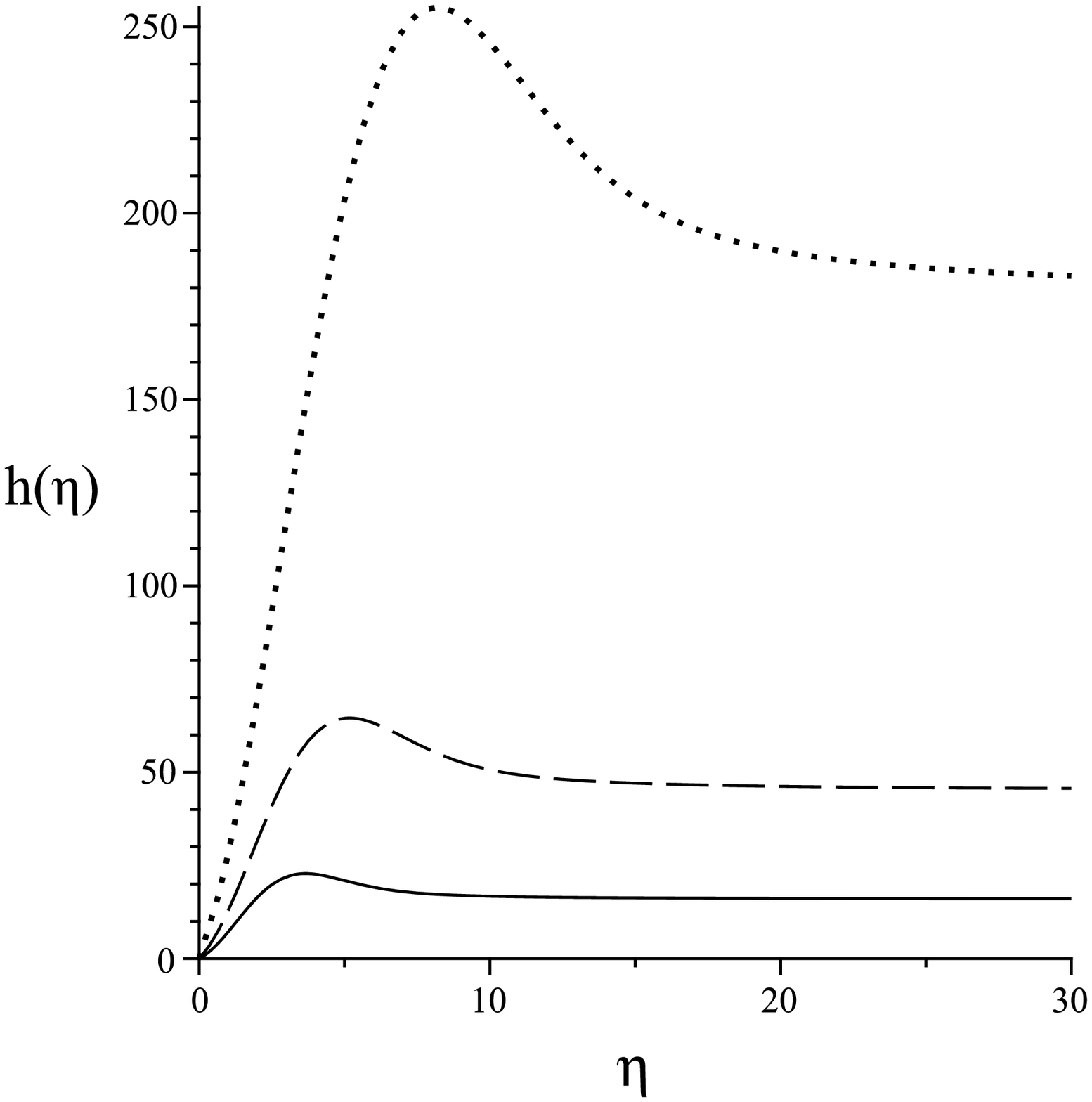}}}
\caption{Different shape functions of the pressure Eq. (12) as a function of $\eta$
for different thermal diffusivity. The integration constants are  taken $  c_1 =  c_2 =1$ 
for all the three curves. We fixed the value of  $eG=1$ as well.  The solid, the dashed and the dotted lines are for $\kappa =1,2,5$ numerical values, respectively.}	
\label{egyes}       
\end{figure}


where the new variable is  $\eta = (x+z)/t^{\beta}$. 
All the five exponents $\alpha,\beta,\delta,\epsilon,\omega $ are real numbers.  (Solutions with 
integer exponents are  the  self-similar solutions of the first kind and sometimes can be obtained from dimensional considerations.) The $f,g,h,l$ objects are called the shape functions of the corresponding dynamical variables. 

After some algebraic manipulations the following constrains are fixed among the self-similarity exponents : $\alpha = \delta = \beta=1/2$, $\epsilon = 1$ and $\omega = 3/2$  which are called the universality relations.  At this point it is worth to mention that now all the exponents have a fix numerical value which simplifies the structure of the solutions.  There is no free exponential parameter in the original  dynamical system,  like an exponent in the equation of state. As an example we mention one of our former study where the compressible NS equation was investigated \cite{imre4} with a free parameter which described different materials.  
 
 These universality relations dictate the corresponding coupled ODE system which has the following form of 

\begin{eqnarray}
-\frac{f}{2} -\frac{f'\eta}{2} + ff' + gf' + h' -2\nu f''  &=& 0, \nonumber \\ 
-\frac{g}{2} -\frac{g'\eta}{2} + fg' + gg' + h' - eGl -2\nu g'' & =& 0, \nonumber \\ 
-\frac{3l}{2} -\frac{l'\eta}{2} + fl' + gl' -2\kappa l'' &=& 0,	  \nonumber \\
f' + g' &=& 0.
\label{ode1}
\end{eqnarray}

From the last (continuity) equation we automatically get the $
f + g = c $  and $ f''+ g'' =0 $ conditions which are necessary in the following. 

Going through a straightforward derivation  the next  single ODE for the shape function of the  temperature distribution can be separated 
\eq
2\kappa l'' + \frac{l'\eta}{2}  + \frac{3l}{2} = 0.
\eqe
The solution is 
\eq
l = c_1\left[ 4 erfi \left( \frac{\sqrt{2}\eta}{4 \sqrt{\kappa}} \right)  \sqrt{2\pi}\left( \kappa - \frac{\eta^2}{4} \right)
e^{-\frac{\eta^2}{8\kappa}}   + 4{\sqrt{\kappa}} \eta \right] + c_2 e^{-\frac{\eta^2}{8\kappa}}(4\kappa - \eta^2) 
\label{temp}
\eqe
where $c_1, c_2$ are free integration constants. The erfi means the imaginary error function 
defined via the integral $ 2/\sqrt{\pi}  \int_0^x  exp(x^2)  dx $ for more  
details see \cite{abr}. 
 It is interesting, that the temperature distribution is separated from the other three dynamical variables an does not depend on the 
viscosity coefficients as well. We may say, that among the solution obtained from the self-similar Ansazt the temperature has the highest priority and this quantity defines the pressure and the velocity field. That is a remarkable feature. 
In a former study, where the one-dimensional Euler system was investigated with heat conduction \cite{imre3}
we found the opposite property, the density and the velocity field were much simpler than the temperature field.  
Figure 2  presents different shape functions of the temperature for different thermal diffusivity values. 
The first message is clear, the larger the thermal diffusivity the larger the shape function of the temperature distribution. 
A detailed analysis of Eq. (9) shows that for any reasonable $\kappa$ and c values the main property of the function is not changing - has one global maximum and minimum with a strong decay for large $\eta$s. A second remarkable feature is the single oscillation which is not a typical behavior for self-similar solutions. We investigated numerous non-linear PDE systems till 
today \cite{imre2,imre4,imre3} some of them are even not hydrodynamical \cite{imre5} and never 
found such a property. This analysis clearly shows that at least the temperature distribution in this physical  system has a 
single-period anharmonic oscillation. 
For a fixed time value and a well-chosen z the difference of values of $\eta$ where $l(\eta)$ yields a minima and a maxima
correspondsto that $ \Delta x $ at which the temperature  (and density) fluctuation may start the Behnard convection. 
To go a step further we may calculate the Fourier transform of the shape function,  $l(\eta)$ 
Eq. (\ref{temp}) to study the spectral distribution. (An analytic expression for the Fourier transform is available, which we skip now.) 
The first term (which is proportion to $c_1$) becomes a complex function, however the general overall shape remains the same, 
a single-period anharmonic oscillation with a global minimum and maximum like on Figure 2.  Of course, the zero transition of the function depends on the value of $\kappa$. 
The second term of the Fourier transformed function which is proportional to $c_2$ remains a Gaussian which is not
interesting.

For completeness we give the full two dimensional temperature field as follows
\begin{eqnarray}
T_1(x,z,t) =&c_1&t^{-3/2}\left[ 4 erfi \left( \frac{ x+z}{4 (\kappa t)^{1/2}} \right) 
 \sqrt{2\pi}\left( \kappa - \frac{(x+z)^2}{4t} \right)
e^{-\frac{(x+z)^2}{8\kappa t}}   +   \frac{4 \sqrt{\kappa} (x+z)}{t^{1/2}} \right] +  \nonumber \\ 
&c_2& t^{-3/2}e^{-\frac{(x+z)^2}{8\kappa t}} \left(4\kappa - \frac{(x+z)^2}{t} \right). 
\end{eqnarray}
The shape function of the pressure field can be obtained from the temperature shape function via the following  equation
\eq
h' = \frac{eGl}{2}
\eqe
with a similar solution to (\ref{temp}) 
\eq
h =  c_1 \left[   2\kappa \sqrt{2\pi} eG  \cdot erfi \left( \frac{\sqrt{2}\eta}{4 \sqrt{\kappa}} \right) \eta e^{-\frac{\eta^2}{8\kappa}}  \right] +  c_2 2eG \kappa \eta  e^{-\frac{\eta^2}{8\kappa}} +  c_3,	
\eqe
this can be understood that the derivative of the pressure is proportional to the temperature.
With the known numerical value of the exponent $\epsilon =1$ the scaled pressure field can be expressed as well   
$ P(x,z,t) = t^{-1} h([x+z]/t^{-1/2}) $. Note, the difference between the $\omega$ and the $\epsilon$ exponents, which are responsible for the different asymptotic decays. 
The temperature field  has a stronger damping for large $\eta$ than the pressure field. 
(It is worth to mention that for the three dimensional NS equation, without any heat exchange 
the decay exponent of the pressure term is also different to the velocity field \cite{imre2}.  ) 

At last the ODE  for the shape function of the velocity component z reads 
\eq
4\nu g'' + g'\eta  + g+eGl = 0 
\eqe
 which directly depends on the temperature on $l(\eta)$ and all the physical parameters 
$\nu, e, G, \kappa $, of course. 
In contrast to the pressure and temperature field there is no closed solutions available for 
a general parameter set.  The formal, most general solution is 
\eq 
g = \tilde{c}_2 e^{-\frac{\eta^2}{8 \nu}} + e^{-\frac{\eta^2}{8 \nu}}  \left\lbrace 
\int \frac{1}{4\nu} \left[   \left(  \tilde{c}_1 - 4 eG \kappa c_2 \eta e^{-\frac{\eta^2}{8\kappa}} 
-4eG\kappa\sqrt{2\pi} c_1 erfi \left(  \frac{\sqrt{2}\eta}{4\sqrt{\kappa}} \right) \eta e^{-\frac{\eta^2}{8\kappa}}  
\right)   e^{\frac{\eta^2}{8\nu}}    \right] d\eta  \right\rbrace
\eqe 
where $\tilde{c}_1$ and $\tilde{c}_2$ are the recent integration constants. 
Note, that the integral can be analytically evaluated if and only if $\nu = \kappa$ which is a great restriction to the physical system. 
We skip this solution now. 
The other way is to fix $c_1 =0$ and let $\kappa$ and $\mu$ free. 
The solution has the next form of 
\eq
g =  \tilde{c}_1 e^{-\frac{\eta^2}{8 \nu}} erf \left(  \frac{\eta}{4} \sqrt{-\frac{2}{\nu}  } \right) 1 + \tilde{c}_2 e^{-\frac{\eta^2}{8 \nu}}   - 
\frac{ 4eGc_2\kappa^2   e^{-\frac{\eta^2}{8 \kappa}}  }{\kappa-\nu}.
\label{geta}
\eqe  
Note, that now the $\nu \ne \kappa$ condition is obtained. The $\tilde{c}_1$ and $\tilde{c}_2$ are the recent integration constants as above, 
it is interesting that if both of them are set to zero, the solution is still not trivial. For a physical system the kinematic viscosity $\nu > 0$ is always positive, therefore 
in the case of $\tilde{c}_1 \ne 0$  the solution becomes complex. 
Figure 4 shows the shape function of the z velocity component. 
It is clear that the real part is a Gaussian function  and the complex part is a Gaussian distorted with an error function, which is an interesting final result. 
In the literature we can find system which shows similarities like the work of Ernst \cite{ernst} who presented a study where a the asymptotic normalized velocity autocorrelation function calculated 
from the linearized Navier-Stokes equation  has an error function shape.    

\begin{figure} 
\scalebox{0.45}{
\rotatebox{0}{\includegraphics{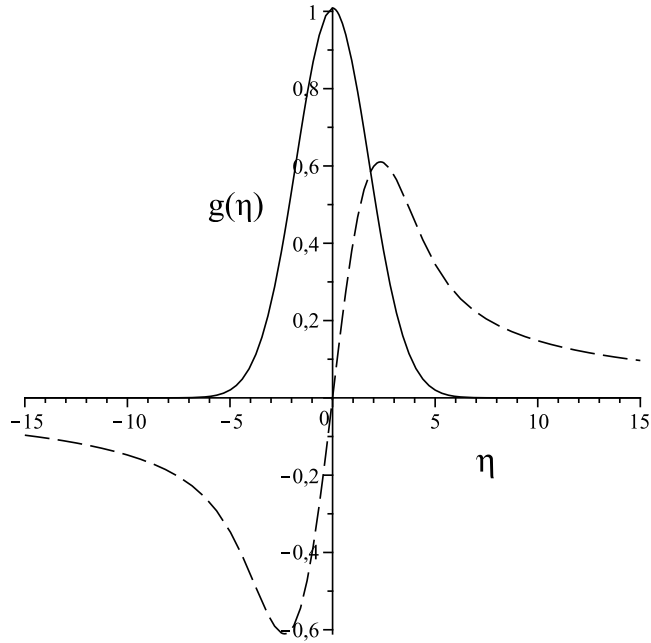}}}
\caption{The shape functions of the z velocity field  $g(\eta)$  Eq. (\ref{geta}) as a function of $\eta$.
The  solid line is the  real and the  dotted is the  complex part.  All the integration constants are  taken $  \tilde{c}_1 =  \tilde{c}_2 = c_2 =1$.  
The physical constant $ eG = 1 $ as well. The $\kappa = 0.04$ and $\nu = 0.8$. }	
\label{negyes}       
\end{figure}


\section{Summary}
We investigated the classical OB equation which is the starting point of countless dynamical and chaotic systems. 
Instead of the usual Fourier truncation method  we applied the two-dimensional generalization of the self-similar Ansatz and found a coupled non-linear ODE system which can be solved with 
quadrature. The main result is that even these kind of solutions - build up from error functions - 
show some oscillating behavior. 
The resulting expressions show the decay of temperature and pressure fluctuations and velocity field in time.   
Due to our knowledge certain parts of the climate models are based on the OB equations therefore our results might be an interesting sign to climate experts.


\section{Acknowledgement}
This work was supported by the Hungarian OKTA NK 101438 Grant. 
We thank for Prof. Barnab\'as Garai for useful discussions and comments.   



\begin{references}

\bibitem{bous} M.J. Boussinesq, 
Rendus Acad. Sci (Paris),  {\bf{72}}, 755 (1871).

\bibitem{ober} A. Oberbeck, Annal. der Phys. und Chemie Neue Folge {\bf{7}}, 271 (1879).

\bibitem{salz} B. Saltzman, J. Atmos. Sci. {\bf{19}}, 329 (1962).  
 
\bibitem{lorenz} E.N. Lorenz, J. Atmos. Sci. {\bf{20}}, 130 (1963) ibid,  {\bf{26}}, 636 (1969) ibid.  {\bf{63}}, 2056 (2005).

\bibitem{tamas} T. T\'el and M. Gruiz, {\it{Chaotic Dynamics}} Cambridge University Press, 2006. 

\bibitem{lain} C. Lainscsek, Chaos {\bf{22}}, 013126 (2012).

\bibitem{curry} J.H. Curry, Commun. Math. Phys. {\bf{60}}, 193 (1978), SIAM J. Math. Anal.
{\bf{10}}, 71 (1979).  


\bibitem{musi} D. Roy and Z.E. Musielak, {\bf{32}}, 1038 (2007) ibid,  {\bf{31}}, 77 (2007), ibid,  {\bf{33}}, 1064 (2007). 

\bibitem{sedov} L. Sedov, {\it{Similarity and Dimensional Methods in Mechanics}} CRC Press 1993. 

\bibitem{barenb} G.I. Baraneblatt, {\it{Similarity, Self-Similarity, and 
Intermediate Asymptotics}} Consultants Bureau, New York 1979. 

\bibitem{zeld} Ya. B. Zel'dovich and Yu. P. Raizer {\it{Physics of Shock 
Waves and High Temperature Hydrodynamic Phenomena}} Academic Press, New York 
1966.

\bibitem{imre2} I.F. Barna, Commun. in Theor. Phys. {\bf{56}}, 745  (2011). 

\bibitem{imre4} I.F. Barna and L.  M\'aty\'as, Fluid. Dyn. Res.  {\bf{46}},  055508 (2014).  

\bibitem{imre3} I.F. Barna and L.  M\'aty\'as, Miskolc. Math. Notes.  {\bf{14}}, 785 (2013).

\bibitem{chandr} S. Chandrasekhar, {\it{Hydrodynamic and Hydrodynamic Stability}}, Chapter II,  Dover, New York 1984. 

\bibitem{berge} P. Berg\'{e}, Y. Pommeau and C. Vidal, {\it{Ordre Within Chaos}},  Appendix D., J. Wiley, 
New York 1984.

\bibitem{spar} C. Sparrow, {\it{The Lorenz Equations: Bifurcations, Chaos, and Strange Attactors}} Springer-Verlag, New York, 1982. 

\bibitem{hilb} R.C. Hilborn, {\it{Chaos and Nonlinear Dynamics}} Appendix C, Oxford University Press 2000. 



\bibitem{abr} M. Abramowitz and I. Stegun, {\it{Handbook of Mathematical Functions}} Dover 
Publication., Inc. New York 1970, Chapter 7, Eq. 7.1.1.                                    

\bibitem{imre5}  I.F. Barna,  Laser. Phys. {\bf{24}}, 086002 (2014).

\bibitem{ernst} M.H. Ernst, E.H. Hauge and J.M.J. van Leeuwen, Phys. Rev. Lett. {\bf{25}}, 1254 (1970). 

\end{references}
\end{document}